\documentclass[pdflatex,sn-mathphys-num]{sn-jnl}

\usepackage{graphicx}
\usepackage{amsmath,amssymb,amsfonts}
\usepackage{booktabs}
\usepackage{tabularx}
\usepackage{array}
\usepackage{xurl}
\usepackage{url}
\usepackage{xcolor}
\usepackage{textcomp}
\usepackage{pifont}
\usepackage{float}
\usepackage{subcaption}
\begin{document}

\title{Quantifying the Impact of Stealthy BLE Spam \& Flooding Attacks on IoT Environments}

\author*[1]{\fnm{Usman} \sur{Rauf}}\email{urauf@mercy.edu}

\author[1]{\fnm{Adalynn} \sur{Martinez}}\email{amartinez@mercy.edu}

\author[2]{\fnm{Fadi} \sur{Mohsen}}\email{f.f.m.mohsen@rug.nl}

\affil*[1]{\orgdiv{Department of Computer Science}, \orgname{Mercy University}, \city{Dobbs Ferry}, \state{NY}, \country{USA}}

\affil[2]{\orgname{Bernoulli Institute for Mathematics, Computer Science and Artificial Intelligence,University of Groningen}, \city{Groningen}, \country{The Netherlands}}

\abstract{

The energy-efficient design of the BLE protocol, emphasis on rapid, and user-friendly discovery, making it an ideal choice for IoMTs, specifically, military field medical systems, and battlefield wearable sensors. Especially in active conflict zones, when static medical facilities are vulnerable and often targeted, limiting their viability for sustained care delivery. 
This rapid deployment, and ease of management comes at the cost of expanded attack surface, i.e., BLE flooding attacks. During such attacks, adversaries flood advertisement channels with unauthorized connection or advertising requests to exhaust nearby device resources and disrupt legitimate communication, sometimes culminating in denial-of-service conditions. A first public proof-of-concept of such attacks, using a Raspberry Pi has since been adapted to commodity platforms (e.g., Flipper Zero, HackRF, Android), lowering the barrier to attack. In contested environments, such platforms are directly relevant to adversarial RF jamming and spoofing operations, where low-cost, portable devices can induce disproportionate disruption in dense wireless ecosystems. In this work, we develop a quantitative foundation for understanding the impact of such attacks and propose a practical deterrence strategy based on agility to raise the cost of such attacks. Our main contributions include:  (1) formalization of this problem as a finite resource stochastic system with dynamic adversarial capabilities under diverse scenarios (e.g., single/multiple \& static/agile attackers), identifying bounds where disruption becomes likely. (2) We also synthesize agility parameters for adaptive deterrence that can lead to the development of actionable guidance for implementers and standards bodies (Bluetooth Interest Group) seeking to harden BLE stacks and preserve the reliability of dense IoT ecosystems. (3) Our third contribution is the cross-validation of our proposed stochastic model with a real-life test bed by implementing several attack scenarios. (4) Our last and main contribution includes the development of integrated agility based framework to neutralize BLE based threat.

}


\maketitle

\section{Introduction}

Bluetooth Low Energy (BLE) protocol has become a cornerstone technology for modern wireless communication, enabling connectivity across a wide range of applications, including personal gadgets, smart home systems, industrial automation, and healthcare. Its widespread adoption is due to its energy efficiency, ease of use, quick deployment, and rapid device discovery. Though the emphasis on ease of use and quick deployment has inadvertently expanded the attack surface, exposing vulnerabilities that adversaries can exploit. For example, BLE's minimum energy consumption and rapid device discovery are achieved via the reliance on lightweight advertising exchanges across three dedicated channels (37, 38, 39). This architecture proved effective for connecting wearables and peripherals, but has also left discovery and advertising as relatively unprotected operations \cite{CHO201672}. Another example is the use of BLE-enabled sensors in healthcare. Remote monitoring systems rely on these sensors to deliver continuous and reliable updates, critical to relay vital signs and trigger timely responses\cite{Chaari2020RHMS}. 
Although these BLE-enabled sensors are vulnerable to adversarial manipulation, security surveys highlight that attacks such as energy drain and denial of service, targeting advertising or mesh communication, can greatly impair the reliability of these devices. This raises concerns regarding clinical safety and monitoring in dense IoMT environments\cite{LACAVA2022108953}.

In military and contested operational environments, the role of BLE-enabled IoMT systems becomes even more critical. Battlefield medical support increasingly relies on wearable sensors, portable monitoring systems, and ad hoc telemedicine platforms to provide real-time patient data in the absence of stable infrastructure. Unlike traditional healthcare settings, these systems operate in dynamic and adversarial RF environments where communication channels are inherently exposed. Static medical facilities and centralized infrastructure may be infeasible due to their susceptibility to targeting, necessitating decentralized and wireless-dependent solutions\cite{who_ssa}. This reliance on BLE introduces a unique vulnerability: the same lightweight, unauthenticated advertising mechanisms that enable rapid deployment also expose these systems to disruption through low-cost RF interference, jamming, and spoofing. As a result, even resource-constrained adversaries equipped with commodity platforms can degrade or deny critical medical communication, amplifying the operational impact of BLE-based attacks beyond conventional IoT scenarios.

An emerging threat in this landscape is the BLE spam/flooding attack, which leverages BLE advertisement channels to flood nearby devices with unauthorized connection requests. These attacks can disrupt normal device operations, degrade user experience, and, in severe cases, lead to denial-of-service (DoS) conditions \cite{MalwarebytesLabs2023FlipperZeroOnline}. Furthermore, adversaries can manipulate BLE spams to interfere with legitimate communication, exploit device behavior, or facilitate other malicious activities. As the adoption of BLE-enabled devices continues to increase, understanding the feasibility, impact, and mitigation of such attacks is critical to securing the IoT ecosystem.

Despite ongoing research on BLE security, the study of BLE flooding attacks remains in its early stages, with limited assessments of their practical implications. The first proof-of-concept for this attack vector was demonstrated in 2023 using a Raspberry Pi equipped with a Bluetooth adapter and an external antenna \cite{Winder2023iOS17Hack}. Since then, it has been adapted for use on various platforms, including Flipper Zero, HackRF, and even Android applications, making it increasingly accessible to attackers\cite{fdroid-ble-spam-2023,mayhem-wiki-blespam-2024,dankelmann-ble-spam-android-2023}. BLE advertisements, as a fundamental component of device discovery and connectivity, inherently lack authentication or pairing requirements, rendering them particularly susceptible to exploitation. 

To address these gaps, this paper presents a quantitative method to understand the repercussions of such attacks using a stochastic modeling approach. We examine the feasibility and impact of these attacks by evaluating key parameters such as device duty cycles, spam transmission rates, and multiattacker scenarios. Furthermore, we investigate adversarial strategies to maximize disruption and assess mitigation techniques, including dynamic countermeasures where BLE devices adaptively adjust their behavior in response to spam attacks. Finally, we conduct an experimental analysis of real-world BLE spam traffic, providing insights on the characteristics of the attack and potential defense mechanisms. Demonstrating that sustained spam bursts can overwhelm BLE devices and degrade nearby IoT activity, underscoring the practical risks that motivate both our stochastic analysis and the evaluation of adaptive defenses.

\paragraph{Road-map:} 
The remainder of this paper is structured as follows: Section~\ref{sec:relatedwork} reviews related work. Section~\ref{sec:methodology} outlines our formalization and modeling methodology. Section~\ref{sec:feasibility} presents our feasibility and impact analysis of the attack. Section~\ref{sec:empirical} shows the empirical analysis. Section~\ref{sec:defense} presents agility and formal methods based defense framework the neutralize  persistent threats. Finally, we conclude the paper in Section~\ref{sec:conclusion}.

\section{Related Work}
\label{sec:relatedwork}

The research landscape relevant to traffic degradation in Internet of Medical Things (IoMT) networks under adversarial BLE activity can be grouped into three streams: (i) modeling efforts that quantify network-quality deterioration, (ii) public BLE datasets, and (iii) attack-analysis and detection frameworks.While each category contributes distinct insights, we emphasize modeling efforts as the most aligned with our goals of capturing adversarial behavior under dynamic conditions. The other two streams offer useful support material but do not engage directly with attacker–defender modeling. 

To structure this review, we organize prior work according to nine modeling capabilities that help distinguish their relevance to attacker–defender dynamics in BLE environments. \emph{Protocol} specifies the communication layer or network context in which the model operates.  \emph{Attack Type} identifies whether adversarial behavior is captured and specifies the forms represented (e.g., DoS, MITM). \emph{Attacker Profiling} denotes the degree to which adversaries are behaviorally characterized, such as by aggression level or operational strategy. \emph{Behavioral Dynamics} considers how interdependent behaviors evolve, for example, the relationship between spam rate and scanning duty cycles. \emph{Agent Agility} evaluates whether agents (attacker or defender) can adapt strategies in response to environmental conditions. \emph{Impact Estimation} assesses whether the consequences of attacks on system performance or communication quality are explicitly quantified. \emph{Model Validation} denotes the degree to which predicted behavioral patterns are cross-validated through experimental setups or controlled testbeds, thereby linking theoretical models to observed evidence. \emph{Engagement Dynamics} examines the diversity of modeled interaction scenarios, including static–static, agile–static, and agile–agile matchups. Finally, \emph{Focus} indicates the primary analytical or operational goal, such as device localization, threshold-based detection, or behavior-based threat modeling.

\noindent \textbf{Modeling efforts for network-quality deterioration:} Prior work across wireless domains has explored stochastic and queueing-theoretic models to quantify degradation caused by interference or malicious behavior. Table~\ref{tab:ble-framework} presents our analysis of these efforts with respect to the nine modeling capabilities introduced.

\begin{table*}[t]
\centering
\begingroup
\setlength{\tabcolsep}{4pt}
\resizebox{\textwidth}{!}{%
  \begingroup
  \begin{tabular}{
  >{\raggedright\arraybackslash}m{2.5cm} 
  >{\centering\arraybackslash}m{2.5cm} 
  >{\centering\arraybackslash}m{2.5cm} 
  >{\centering\arraybackslash}m{2.5cm} 
  >{\centering\arraybackslash}m{2.5cm} 
  >{\centering\arraybackslash}m{2.5cm} 
  >{\centering\arraybackslash}m{2.5cm}
  >{\centering\arraybackslash}m{2.5cm}
}
\toprule
\textbf{Capability} & \textbf{Halloush \& Liu }~\cite{halloush2019smp} & \textbf{Zhang et al. }~\cite{wang2023hmm} & \textbf{Shanbhag et al.} ~\cite{shanbhag2019mdp}& \textbf{Kavousi Ghafi et al.}~\cite{KavousiGhafi2021} & \textbf{Bhunia et al.}~\cite{Bhunia2014} & \textbf{Allouzi \& Khan}~\cite{IoMT_Edge_Network} & \textbf{Our Behavioral Model} \\
\midrule

\textbf{Protocol} & Smart Grid / Generic Wireless & UAV & Cognitive Radio / WSN & BLE Coexistence (WLAN/BLE) & Cognitive Radio Networks & IoMT Edge Networks & BLE \\

\textbf{Attack Type} & Jamming (DoS) & Stealthy Jamming & Adaptive Jamming & \ding{55} & Jamming (DoS) & DoS, Spoofing, Malware, etc. & DoS, Adaptive Spam \\

\textbf{Attacker Profiling} & Partial\newline (2 operational types) & Partial\newline (Hidden jammed/unjammed states) & \checkmark & \ding{55} & \ding{55} \newline (Beyond paper scope) & \ding{55} & \checkmark \\

\textbf{Behavioral Dynamics} & \checkmark & \checkmark \newline (Timing States) & \checkmark & \ding{55} & \ding{55} \newline (Beyond paper scope) & \ding{55} & \checkmark \\

\textbf{Agent Agility} & Partial \newline (Defender) & \ding{55} \newline (Detection Only) & Partial \newline (adaptive attacker) & \ding{55} & Partial \newline (honeypot defense dynamics) & \ding{55} & \checkmark \\

\textbf{Impact Estimation} & \checkmark  & \ding{55} & Partial \newline (total error probability) & \checkmark \newline (BER/PER under interference) & \checkmark & Partial \newline (threat likelihood distribution) & \checkmark \\

\textbf{Model Validation} & Testbed
cross-validation  & Simulation-based & Simulation-based & IQ \& spectrum traces & Simulation-based & Simulation-based with CVSS probabilities & Testbed cross-validation \\

\textbf{Engagement Dynamics} & Limited \newline (single jammer) & Limited \newline (single link/jammer) & Limited \newline (single attacker; adaptive attacker vs static victim) & \ding{55} & Limited \newline (attacker static) & \ding{55} & \checkmark \\
\textbf{Focus} & QoS under Jamming & Detection of Stealthy Jamming & Adaptive Adversary Strategy & Co-Channel Intelligent Interference & Anti-jamming with Honeypot & Threat Likelihood Estimation & Behavior-based Threat Modeling \\
\bottomrule
\end{tabular}
  \endgroup
}
\caption{Summary of prior modeling efforts across wireless domains versus our BLE spam model, emphasizing outcome granularity and validation approach.}
\label{tab:ble-framework}
\endgroup
\end{table*}

As shown in Table~\ref{tab:ble-framework}, most prior models focus on jamming in various wireless networking contexts.  Yet comparable work for BLE remains sparse. While many identify denial-of-service (DoS) as an attack type, few attempt \emph{attacker profiling} or capture fine-grained \emph{behavioral dynamics}. In particular, attributes such as \emph{agent agility} and \emph{engagement dynamics} are largely missing. Such treatments overlook adaptive adversaries and do not link attacker send rates or duty-cycle exploitation to system-level degradation.

Stochastic abstractions have been employed to model the impact of interference on quality-of-service (QoS) in wireless communication environments. One example, introduced by Halloush and Liu is a semi-Markov process (SMP) to model jamming-resilient systems, enabling derivation of availability metrics and packet loss probabilities under varying levels of adversarial intensity\cite{halloush2019smp}. Similarly, Bhunia et al. introduced a queuing-theoretic model for cognitive radio networks incorporating honeypot-based defenses to analyze the effects of jamming on packet delays and loss\cite{Bhunia2014}. While both models illustrate the feasibility of quantifying long-term service degradation, their analyses remain confined to coarse-grained QoS metrics. They do not directly link interference conditions to outcomes in device discovery or session outcomes.

To capture stealth and adaptivity, learning-based and state-driven approaches extend this line of work. Zhang et al. apply a hidden Markov model (HMM) to detect low-rate jamming in UAV-assisted wireless networks, inferring adversarial behavior from received energy traces\cite{wang2023hmm}. Shanbhag et al. adopt a Markov decision process (MDP) framework to capture adaptive jamming strategies that evolve through environmental feedback\cite{shanbhag2019mdp}. Although these approaches highlight the agility of the attacker and his behavioral dynamics, they primarily report performance in terms of detection accuracy or the efficiency of the attacker's policy, rather than the impact on system-level services.

Efforts more closely aligned with BLE include the work of Kavousi Ghafi et al., who developed measurement-driven interference models using IQ-domain and spectral traces. Their study characterizes packet error rates under co-channel interference from BLE and WLAN sources, demonstrating that interference effects can be captured with high statistical fidelity \cite{KavousiGhafi2021}. However, the scope remains centered on coexistence challenges rather than deliberate adversarial behavior.

Allouzi and Khan contributed a Markov-chain-based modeling framework for IoMT edge networks, estimating attack transition probabilities based on device vulnerabilities and CVSS scores\cite{IoMT_Edge_Network}. This work illustrates how stochastic state transitions can effectively represent diverse attack vectors in medical environments. Yet the analysis is broad, covering spoofing, DoS, and malware, without attention to BLE-specific broadcast channels. Specifically, it focuses on the probability of various threat occurrences rather than assessing the capabilities of the attacker.

\noindent \textbf{Public BLE datasets}: BLE datasets offer valuable empirical support for tasks such as device classification and anomaly detection. However, they offer limited support for adversarial modeling. BLEBeacon \cite{BLEbeacon}, SDR4IoT \cite{sdr4iot}, Bouaru et al. university exam captures \cite{universitydataset}, and CICIoMT2024 \cite{CICIoMT2024} consist primarily of benign traffic. Only minimal adversarial content is observed in CICIoMT2024. Crucial features, such as \emph{attacker profiling}, \emph{agent agility}, and \emph{engagement dynamics}, are either absent or heavily abstracted. BlueTack \cite{bluetack} (see also \cite{abad2022bluetack}) is a partial exception, compared to the previous datasets, for IDS evaluation in IoMT. The dataset centers on DoS and MITM scenarios, but the attack traces are aggregated into fixed time windows, limiting analysis of fine-grained behavioral dynamics. As a result, the data cannot directly capture agent agility or reveal engagement dynamics.

\noindent \textbf{Attack-analysis and detection frameworks:} Protocol and system analyses advance state-aware checking, robustness testing, and target-side safeguards. This includes tools and studies such as BlueSWAT \cite{BlueSwat}, BLESS \cite{BLESS}, BLEdiff \cite{BLEdiff}, formal-model–driven discovery \cite{FormalModel}, and security testing frameworks \cite{securityTestingFramework}. These resources document or emulate \emph{attack types} (e.g., misconfiguration abuse, DoS, or state disruption) and provide \emph{model validation} via testbeds/case studies. However, \emph{attacker profiling} remains limited (focus is device/app/implementation rather than adversary identity/rate control), \emph{behavioral dynamics} are only partially captured, \emph{agent agility} is rarely parameterized (attackers/defenders typically static), and \emph{engagement dynamics} (coordination among multiple attackers, or closed-loop A$\leftrightarrow$D
coupling) are generally out of scope. Domain-specific IoMT defenses (e.g., MITM detection for eHealthcare BLE \cite{Marc}) further motivate adversary-aware modeling but similarly treat the attacker as largely static. Complementary attack studies such as demonstrations of BLE DoS feasibility \cite{gullberg2017ble_dos,ditton2020ble_pocdos}, protocol-level manipulations \cite{castro2019ble_injection}, and the SweynTooth family of implementation faults \cite{garbelini2020sweyntooth}, clearly establish \emph{attack type} coverage. Although they offer empirical validation, they typically omit \emph{attacker profiling}, do not model \emph{behavioral dynamics} beyond simple floods, and do not consider \emph{agent agility} or \emph{engagement dynamics}.

\textbf{Gap and our contribution}
Across these streams, existing work rarely (a) profiles the \emph{attacker} with per-attacker timing/rate parameters, (b) models \emph{behavioral dynamics} at the advertisement timescale, (c) exposes \emph{agent agility} on both sides, or (d) explicitly models \emph{engagement dynamics} (e.g distinguishing between independent and coordinated attackers, as well as non-adaptive (open-loop) and adaptive (closed-loop) attacker–defender interactions) and links these dynamics to measurable traffic degradation in IoMT environments.. Our probabilistic approach addresses this gap by (1) defining per-attacker success over BLE duty cycle and send-rate, (2) extending to the multi-attacker case via the complement-of-no-success formulation, (3) allowing attacker/defender agility as scenario parameters, and (4) validating alignment with a controlled testbed to link model predictions to observed packet distributions and drop rates.

\section{Problem Description \& Formalization}
\label{sec:methodology}
Calculating the probability of a successful attack on a device operating on low-energy protocol in this scenario involves considering the timing and the probability that the device will be active and able to receive flood/spam messages during its active listening periods. Assuming that a BLE device actively listens for a random amount of time (\textbf{\textit{Active Listening Time}})  before backing-off, or going to sleep mode (\textbf{\textit{Sleep-Time}}), and cannot receive messages in sleep mode, the success probability of an attacker is highly dependent on the availability of the channel opening, and its capability to Spam/flood it. Hence, using these parameters, the problem of estimating attack impact or success probability of flooding/spamming a victim can be modeled as follows:
\begin{itemize}
\item \textbf {Active Listening Time (ALT):} 
We represent the window of average time the BLE device actively listens before going to sleep mode as active listening time (ALT).
Whereas, \textbf{\textit{ALT}} $\in \mathbb{R}^{+}$ (measured in milliseconds or seconds).\\

\item \textbf{Sleep Time (ST):} 
This variable represents the window of time in which a device chooses not to actively participate in receiving certain types of traffic, in this case, unauthorised advertisements. This notion is very similar to the back-off mechanism in Carrier Sense Multiple Access (CSMA CA/CD), but instead of activating/assigning a back-off window after sensing the channel, this sleep time is determined solely by the device's preference and risk tolerance, without scanning any channels. Whereas, \textbf{\textit{ST}} $\in \mathbb{R}^{+}$ (measured in milliseconds or seconds).
\\
\item \textbf{Duty Cycle (DC):} 
The duty cycle is the ratio of the active listening time to the total cycle time (sum of active listening time and sleep time).

\[
DC = \frac{ALT}{ALT + ST}, \quad \text{where } DC \in [0,1] \subset \mathbb{R}
\]

\item \textbf{Spam Rate (SR):} 
The spam rate represents the frequency at which an attacker sends spam messages over time. It is inversely proportional to the interval between consecutive spam attempts, meaning the shorter the interval between messages, the higher the spam rate. Mathematically, this relationship can be expressed as:

\[
SR = \frac{1}{T_{SA}}, \quad \text{where } SR \in \mathbb{R}^{+}, \; T_{SA} \in \mathbb{R}^{+}
\]

Here, $T_{SA}$ denotes the \textit{\textbf{time between spam attempts}}, measured in seconds or milliseconds, and \textbf{SR} denotes the \textit{\textbf{spam rate}}, measured in messages per unit time (e.g., per second). This relationship reflects the standard rate–period dependency, where both variables are strictly positive real numbers. For example, if an attacker sends one spam message every $20$ milliseconds ($T_{SA}=0.02\,s$), the corresponding spam rate would be $SR=\frac{1}{0.02}=50$ messages per second.

\item \textbf{Probability of Spam Success (PS):} \\
Let \(\Delta t\) (in milliseconds) denote the \textbf{\emph{observation window}} during which a single
spam packet is transmitted.

Then given \textbf{\textit{DC}}, and \textbf{\textit{SR}}, the probability of successfully delivering a spam can be formally defined as a product of independently transmitting a spam at a rate (\textit{\textbf{SR}}), while listener is active with a specific duty cycle (\textit{\textbf{DC}}), within the same observational window ($\Delta(t)$). Therefore, the formal expression becomes:
\[
P_{Spam} = DC \times SR \times \Delta t, \quad \text{where } P_{Spam} \in [0,1] \subset \mathbb{R}
\]

For example, if the attacker sends spam every 20 milliseconds (\textit{SR=} $\frac{1}{20 \text{ ms}}$) and the duty cycle is 10\% (\textit{DC=0.1}), then the probability of success would be $P_{Spam} = 0.1 \times \frac{1}{20 \text{ ms}} \times 20 \text{ } ms $. Thus there is a \(10\%\) chance that this particular packet will be received. The same framework generalises to many packets or many attackers by the
complement–probability expressions introduced later.
\end{itemize}

It's important to note that these calculations provide an estimate, and the actual success probability may vary based on the specific behavior of the BLE device and the attacker's strategy. Therefore, to understand the impact of such attacks and their likelihood, a more robust model needs to be formalized, which can take into account multiple attackers with static and dynamic attack parameters. With this objective in hand, our next objective is to develop a formalization to measure the impact of multiple attackers. 

The probability of successfully spamming a target in a scenario where there are multiple attackers can be calculated using the complement probability theorem. For this the base case becomes the scenario in which at least one attacker is successful in conducting the spam attack. The probability that at least one of the attackers succeeds can be calculated as follows:
\begin{equation}
P_{\text{Spam}}(\text{AOS}) = 1 - P_{\text{Spam}}(\text{NS})
\label{eq:at_least_one_success}
\end{equation}
Where: Term $P_{Spam}(AOS)$ refers to the probability of at least one attacker succeeding, and term $P_{Spam}(NS)$ refers to the probability of no one (attacker) succeeding.

If there are N attackers, and $P_{Spam}^i$ represents the probability of success for $i_{th}$ attacker, then the probability of no success for any attacker can be calculated as the product of probabilities of no success for each attacker:
\begin{equation}
P_{\text{Spam}}(\text{NS}) = \prod_{i=1}^{N} (1 - P_{\text{Spam}}^i)
\label{eq:no_attacker_success}
\end{equation}
Therefore, the final expression for the probability of at least one attacker succeeding in a spam attack becomes:

\begin{equation}
P_{\text{Spam}}(\text{AOS}) = 1 - \prod_{i=1}^{N} (1 - P_{\text{Spam}}^i)
\label{eq:final_spam_success}
\end{equation}

We use the above-mentioned formalization (for $P_{Spam}(Atleast-One-Succeeds)$) as our new base model to analyze the attack parameters in details, and build two different variations on it under attack models in the forthcoming section.

\section{Feasibility of Attack \& Impact Analysis}
\label{sec:feasibility}
As our main objective is to design a deterrence strategy for flooding/spam attacks, the very first step towards this direction is the understanding of "\textit{how deadly this attack can be}", and whether it is even possible to design a deterrence against such attacks by altering underlying parameters, i.e., duty cycle, and back-of-window. 

Understanding the feasibility of such attacks on low-energy protocols requires a structured analysis of how different attack models influence the probability of a successful spam/flooding attempt. In this section, we mathematically formalize attack models and analyze these models under different attack scenarios that vary in terms of attacker and defender capabilities and strategies. The goal is to quantify the likelihood of success under different assumptions and to highlight the impact of attacker adaptability on infrastructure security, and eventually devise a strategy to counter these attacks. 
\subsection{Attack Models}
\begin{figure}[h]
    \centering
\includegraphics[width=.75\linewidth]{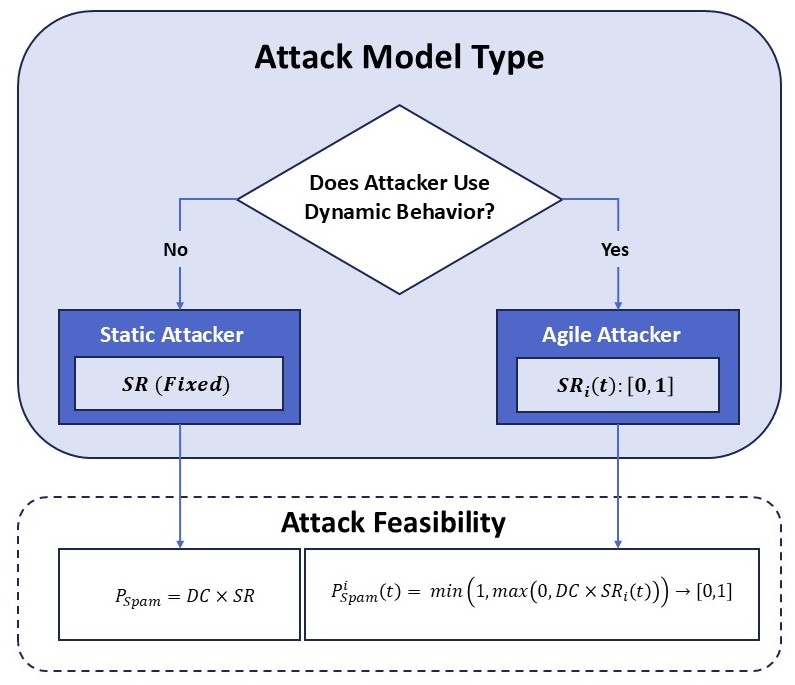}
    \caption{Conceptual overview of the two attack model types considered in this work. A static attacker uses a fixed spam rate, whereas an agile attacker varies its transmission behavior over time. In both cases, the feasibility of a successful flooding attempt is determined by the interaction between the attacker’s spam rate and the victim’s duty cycle. }
    \label{fig:attackModel_Fig}
\end{figure}
Towards this direction, we abstract the adversary’s behavior into two canonical cases that describe the threat landscape. \textbf{\textit{The Static Model}}, that embodies a memory-less attacker with a fixed transmission pattern, energy budget, and consistency; this permits closed-form insight into how a victim’s duty-cycle parameters alone govern vulnerability. In contrast, the \textbf{\textit{Agile Model}} represents an adaptive adversary that continuously changes its emission schedule in response to observed channel opportunities, thereby exploiting any temporal slack in the protocol. The side-by-side development and Study of these models (c.f. Fig. \ref{fig:attackModel_Fig} ) clarifies the security margin lost once an attacker gains even limited situational awareness and sets the foundation for defence strategies that remain effective across both extremes.
\subsubsection{\textbf{Static Model: Single Static Attacker Vs. Static Victims}}

The static model represents the simplest case of spam attacks, where a single attacker uses a fixed strategy against victims whose duty cycles remain constant. To establish a baseline, we first analyze the one-to-one scenario (a single attacker against a single victim). This allows us to observe how the victim’s duty cycle could potentially influences the probability of a successful spam attempt.  

\begin{figure*}[!t]
\centering

\begin{subfigure}{0.49\textwidth}
    \centering
    \includegraphics[width=\linewidth]{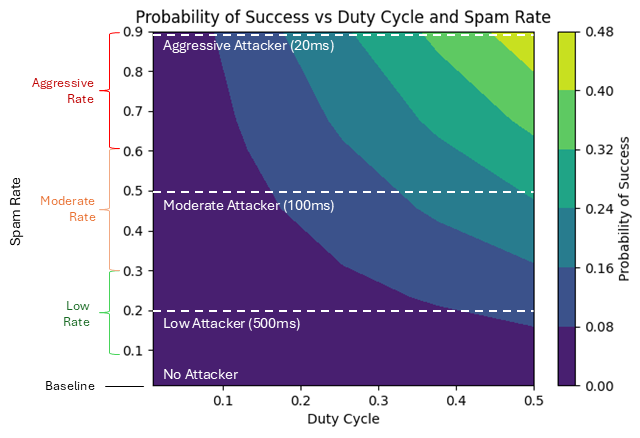}
    \caption{Spam success probability across $DC\in[0.01, 0.5]$.}
    \label{fig:static_low_dc}
\end{subfigure}\hfill
\begin{subfigure}{0.49\textwidth}
    \centering
    \includegraphics[width=\linewidth]{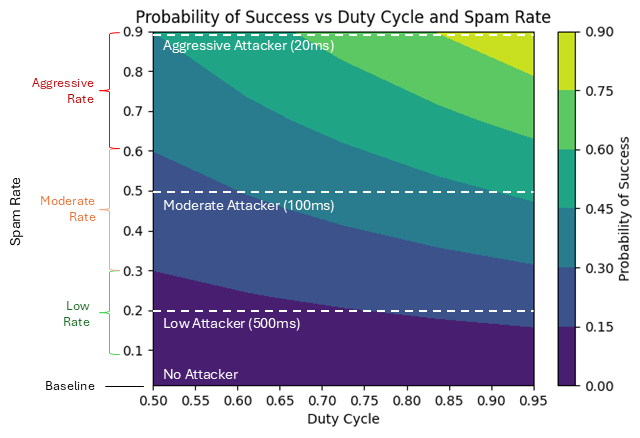}
    \caption{Spam success probability across $DC \in [0.5, 0.9]$.}
    \label{fig:static_high_dc}
\end{subfigure}

\caption{Duty cycle vs. spam success probability ($P_{\mathrm{Spam}}$).}
\label{fig:probability_calculation}
\end{figure*}

Using the base model, we vary the victim’s duty cycle across two ranges: a low range ($DC \in [0.01, 0.5]$) and a high range ($DC \in [0.5, 0.9]$). Figures~\ref{fig:static_low_dc} and~\ref{fig:static_high_dc} show that higher duty cycles correlate strongly with higher spam success probability, confirming that victims with longer active listening periods are more vulnerable to static attackers.  

While this experiment considers a one-to-one case, the insights generalize to scenarios with multiple static victims, as each victim’s probability of being successfully spammed depends primarily on its own duty cycle and the attacker’s fixed spam rate. This baseline result will serve as a reference for more complex models in the following subsections.
This baseline case highlights how static assumptions simplify the analysis but also limit the attacker’s effectiveness. In the next subsection, we extend the model to consider \textit{agile attackers}, who adapt their strategies dynamically, significantly increasing the impact on victims.

To systematically evaluate attacker behavior across varying transmission intensities, we define three attacker profiles based on advertisement interval selection. Under normal conditions, fast BLE advertising typically operates in the range of 100 to 500 ms \cite{argenox_ble_advertising_primer}. This serves as a reference baseline for attacker profile classification.

We define an aggressive attacker transmitting at intervals below 100 ms, exceeding conventional fast-advertising rates and increasing channel saturation. A moderate attacker operates within the 100–500 ms range, aligning disruptive transmissions with normal traffic patterns to improve stealth. A low-rate attacker transmits at intervals above 500 ms, reducing observable deviation from baseline behavior while sacrificing immediate spam impact. While the static model assumes a fixed spam rate, this classification enables systematic evaluation of how varying transmission intervals influence detectability and victim success probability.

\subsubsection{\textbf{Agile Attack Model: Multiple Agile Attacker Vs. Static Victim}}
Unlike the static case, agile attackers adapt their behavior dynamically to maximize impact. They can adjust their spam rate (stochastic) over time, making their attacks more effective and harder to defend against. 

To understand and model the impact of an agile attacker, we must revisit our base model for multiple attackers. The terms $P_{Spam}(AOS)$ and $P_{Spam}(NS)$ in the base model, will no longer be static, and deterministic now due to the involvement of a random variate $SR_{i}(t):\mapsto{[0,1]}$, which represents spam rate of $i^{th}$ attacker at an arbitrary time/iteration (t). Therefore, the final expression in the base scenario ($P_{Spam}^{AOS}$) updates to $P_{Spam}^{AOS} (t)$, and the resultant model, which encapsulates the attacker's agility, can be formally written as follows:

\begin{equation}
P_{\text{Spam}}{\text{NS}}(t) = \prod_{i=1}^{N} \left(1 - P^i_{\text{Spam}}(t)\right)
\label{eq:agile_no_success}
\end{equation}
\textbf{Whereas:} 
\begin{equation}
P^i_{\text{Spam}}(t) = \min\left(1, \max\left(0, DC \times SR_i(t)\right)\right) \mapsto [0,1]
\label{eq:agile_individual}
\end{equation}

\begin{equation}
P_{\text{Spam}}^{\text{AOS}}(t) = 1 - P_{\text{Spam}}^{\text{NS}}(t) = 1 - \prod_{i=1}^{N} \left(1 - P^i_{\text{Spam}}(t)\right) \mapsto [0,1]
\label{eq:agile_final}
\end{equation}

\begin{figure}[!t]
\centering

\begin{subfigure}{0.48\columnwidth}
    \centering
    \includegraphics[width=\linewidth]{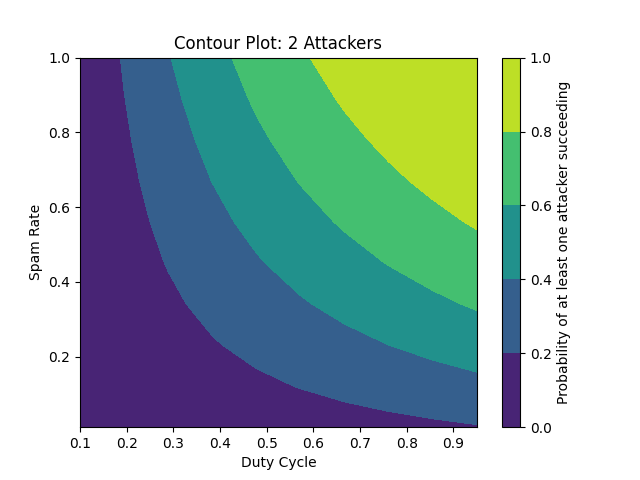}
    \caption{2 agile attackers}
    \label{fig:2agile}
\end{subfigure}
\hfill
\begin{subfigure}{0.48\columnwidth}
    \centering
    \includegraphics[width=\linewidth]{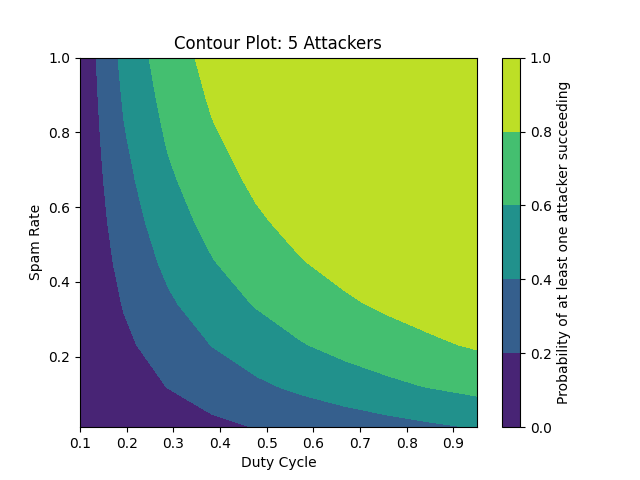}
    \caption{5 agile attackers}
    \label{fig:5agile}
\end{subfigure}

\vspace{0.5em}

\begin{subfigure}{0.65\columnwidth}
    \centering
    \includegraphics[width=\linewidth]{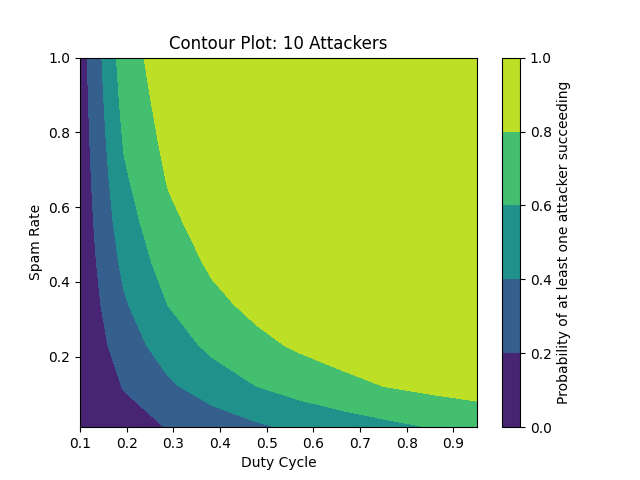}
    \caption{10 agile attackers}
    \label{fig:10agile}
\end{subfigure}

\caption{Impact of agile attackers on attack success rate ($SP_{\mathrm{Spam}}^{\mathrm{AOS}}(t)$) over time.}
\label{fig:multi_agile_attackers}

\end{figure}

\section{Empirical Analysis and Comparison}
\label{sec:empirical}
\subsection{Testbed Setup}
\begin{figure*}[!t]
    \centering
    \includegraphics[width=0.75\textwidth]{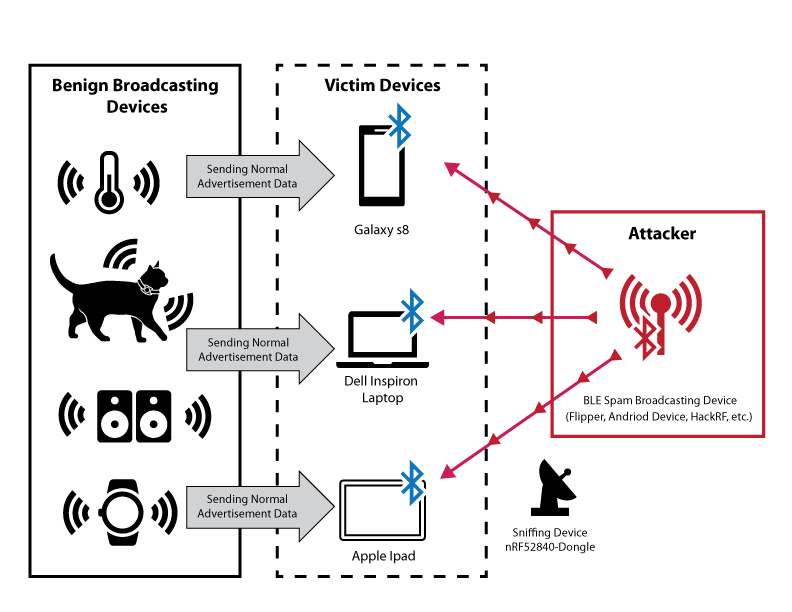}
    \caption{The testbed emulates realistic BLE communication with benign broadcasting devices (thermometer, smartwatch, AirTags) and three victim hosts, including a Galaxy S8 (Android), Dell Inspiron (Windows), and iPad (iOS). A BLE spam transmitter (e.g., Flipper Zero) generates adversarial traffic, while an nRF52840 dongle captures packets for analysis.}
    \label{fig:testbed}
\end{figure*}

Testbeds are designed to mimic various real-world environments where BLE devices operate, such as indoor, outdoor, and different interference scenarios. This includes setting up barriers, varying distances between devices, and introducing controlled interference to simulate a wide range of conditions. For this test, we will initially use low-cost market equipments, as described in Figure \ref{fig:testbed}. Specifically, we construct the testbed around four types of cheap or readily available equipments: BLE sensors, BLE host devices, an attacker device responsible for generating adversarial behavior, and a BLE sniffer for monitoring. 

As shown in Figure \ref{fig:testbed}, benign BLE devices continuously broadcast advertisement traffic toward nearby host platforms while sharing the same wireless space with the attacker device. The attacker operates as an external broadcaster capable of injecting adversarial advertisement packets without requiring prior pairing or authentication, reflecting the opportunistic nature of BLE spam attacks. 

\subsubsection{Required Hardware}

\textit{\textbf{BLE Sensors:}} These could be any BLE sensor, which tends to broadcast readings/measurements taken from its surroundings. Our experimental testbed incorporates a diverse range of BLE–enabled sensors and user devices to emulate the heterogeneity found in real-world environments. The setup includes a Govee Smart Thermo-Hygrometer H5075 for environmental monitoring, a Samsung Galaxy Watch 4 representing wearable systems, and two mobile Apple AirTags attached to pet collars to simulate dynamic movement within the test area. Operating in an open environment within a small office, the testbed continuously captures ambient Bluetooth activity from surrounding devices entering and exiting the vicinity. This background activity introduces realistic environmental noise from various Bluetooth and BLE peripherals such as headsets, speakers, and other transient consumer devices. This provides a more authentic representation of wireless interference in a given environment.

\textit{\textbf{Victim Devices:}}
BLE sensors are not self-sufficient, which means we need a host to control or extract data from them, in IoMT it could be any client-tailored/custom upgraded smart tablet, whereas outside of IoMT settings, the monitoring equipment is usually smartphones, tablets, and/or computers. For broad coverage, our testbed includes one representative host from each category, each selected to reflect a different commercial BLE host ecosystem and common user platform. The chosen hosts act as victim devices that communicate with nearby BLE sensors: a Samsung Galaxy S8 smartphone running Android (representing mainstream Android phones), a ninth-generation Apple iPad (representing iOS/tablet clients), and a Dell Inspiron 14 5425 laptop running Windows 11 Pro (64-bit), equipped with an AMD Ryzen 7 5825U processor, 16 GB DDR4 memory, and a 512 GB NVMe SSD storage device (representing Windows-based computer hosts). Together, these devices provide heterogeneous endpoints for evaluating cross-platform BLE interactions and monitoring behavior across typical real-world host environments.

\textbf{\textit{Attacker Device:}} To mimic a persistent attacker, we used a Flipper Zero device with BLE spam installed. Flippers have a tendency to broadcast traffic on multiple frequency channels, by channel and frequency hopping. Users can also randomize the MAC address for each packet sent or change the DoS/spamming speed from 20ms to 500ms based on the victim's device OS. For analysis purposes we will only utilize changing the spam interval rate, as MAC address randomization complicates the identification process.

\textit{\textbf{Sniffing Devices:}}
 For capturing packets on the data link layer and network layer (layer 2), we aim to employ Wireshark with the integration of a nRF52840 dongle flashed with Nordic’s
nRF Sniffer for Bluetooth LE will be attached to a host machine running Wireshark to complete the setup of our testbed. To ensure that packets between the attacker device and the victim devices are captured properly the dongle will be placed between these devices. For the initial test we aim to have the main victim, sensor and attacker devices stationary to collect baseline data. Victim and attacker devices would be kept within a meter radius of the dongle to reduce the risk of lost packets at increase distances.

While this configuration provides an affordable and accessible BLE sniffing solution, it introduces certain operational limitations. The nRF52840 dongle can monitor only one BLE advertisement channel at a time and does not support true simultaneous multi-channel capture. As a result, packet loss may occur, particularly in environments with elevated channel congestion or high advertisement density. In noisier settings, rapid channel hopping and overlapping transmissions can further increase the likelihood of missed packets. Which must be considered when interpreting captured traffic distributions.

\subsection{Evaluation}

To validate the behavioral trends predicted by our probability estimation model in section 4 (Figure~\ref{fig:probability_calculation}), we conducted controlled data collection of BLE packet transmissions under varying attacker conditions. The resulting comparison, summarized in Table~\ref{tab:ble_spam_distribution}, presents the distribution of BLE packets across three scenarios: Baseline (no attacker), Moderate Attacker (100 ms spam interval), and Aggressive Attacker (20 ms spam interval). Each capture session lasted one hour, with random MAC address rotation disabled to ensure consistent filtering of attacker-generated traffic. The primary objective was to assess how varying spam rates influence real-world packet dominance and to evaluate the extent to which these measurements align with the model’s simulated probability estimates.

\begin{table*}[!t]
    \centering
    \caption{Distribution of BLE packets across scenarios with varying spam rates}
    \label{tab:ble_spam_distribution}
    \resizebox{\textwidth}{!}{%
    \begin{tabular}{@{}l l c c c c@{}}
    \toprule
    \textbf{Scenario} & \textbf{Spam Rate} & \textbf{Attacker Packets} & \textbf{Normal Packets} & \textbf{Total Packets} & \textbf{\% Spam} \\
    \midrule
    Baseline & n/a & 0 & 210{,}107 & 210{,}107 & 0.00 \\
    Moderate Attacker & 100 ms & 99{,}644 & 234{,}275 & 333{,}919 & 29.84 \\
    Aggressive Attacker & 20 ms & 450{,}153 & 211{,}720 & 661{,}873 & 68.01 \\
    \bottomrule
    \end{tabular}
    }
\end{table*}

In the Baseline condition, no attacker activity was present. The capture contained 210,107 normal BLE transmissions, corresponding to a measured spam rate of 0\%. As predicted, the model estimated a success probability of zero across all scanning duty cycles, consistent with expectations in an uncompromised environment.

 In the Moderate Attacker condition (100 ms spam interval), the attacker introduced 99,644 additional packets, increasing the total observed traffic to 333,919. This represents an approximate 30\% increase in captured packets from the baseline. Given that the maximum delay setting available on the Flipper device is 500 ms, the 100 ms injection rate corresponds to the middle region of our simulation’s probability curve. For moderate scanning duty cycles, the model predicts a success probability of approximately 15–45\%, which closely matches the empirical results. At this rate, attacker traffic coexists with legitimate BLE communication but begins to noticeably reduce available channel capacity for normal transmissions.

In the Aggressive Attacker condition (20 ms spam interval), the attacker injected 450,153 and comprising approximately 68\% of the total captured traffic. Operating at five times the rate of the 100 ms condition, this configuration aligns with the upper bound of the simulated probability estimates. Even moderately active receivers (duty cycle:  60\%) experienced modeled success probabilities above 50\%, with higher duty cycles approaching near-total disruption. In practice, an attacker operating at a 20 ms interval can inject sufficient traffic to substantially degrade or even suppress BLE communications for devices that rely on opportunistic scanning.

\begin{figure*}[!t]
\centering
\begin{minipage}[t]{0.495\textwidth}
    \centering
    \includegraphics[width=\linewidth]{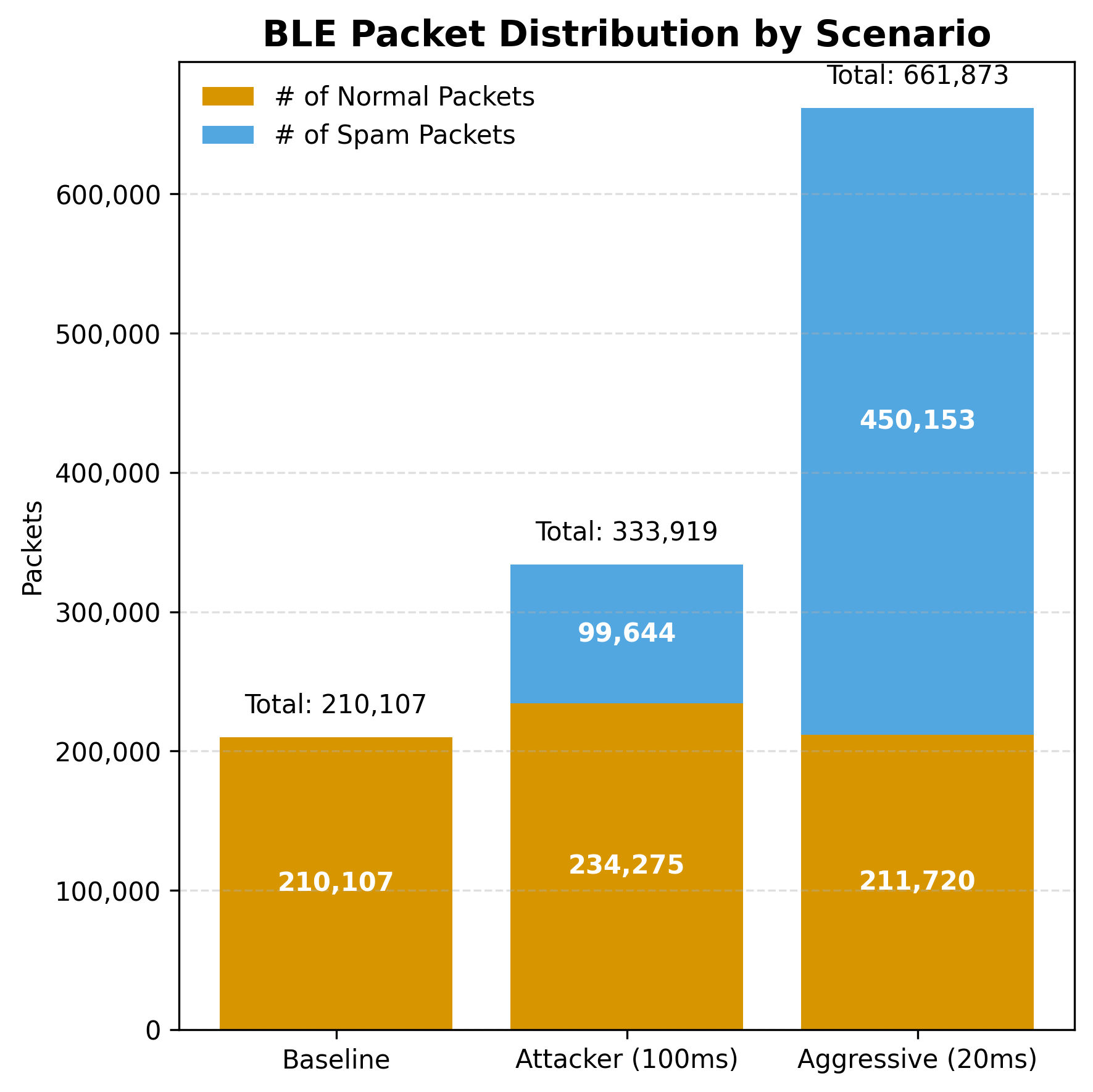}
    \caption{Absolute BLE packet counts across scenarios.}
    \label{fig:packet_dist_stack}
\end{minipage}
\hfill
\begin{minipage}[t]{0.496\textwidth}
    \centering
    \includegraphics[width=\linewidth]{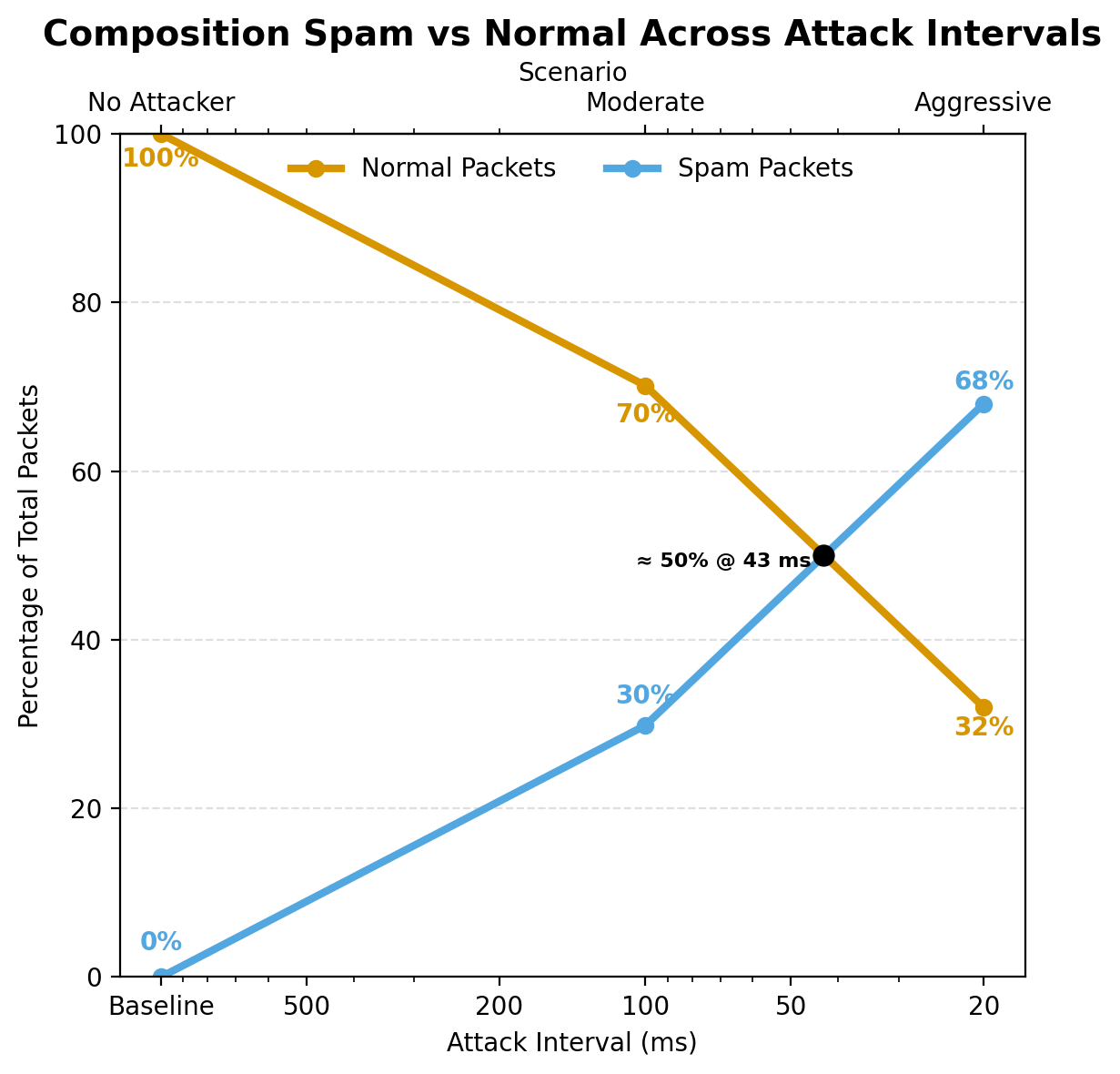}
    \caption{Relative packet composition across attack intervals.}
    \label{fig:packet_dist_slope}
\end{minipage}
\end{figure*}

The absolute distribution of BLE traffic across each attack condition is illustrated in Figure \ref{fig:packet_dist_stack}. Although the volume of legitimate BLE transmissions remains relatively stable across all scenarios, attacker-generated traffic increases substantially as advertisement intervals decrease. In the aggressive 20 ms condition, adversarial packets exceed legitimate traffic by more than a two-to-one ratio. Demonstrating the ability of high-rate spam transmissions to dominate the observable BLE channel. The corresponding packet composition across attack intervals is depicted in Figure \ref{fig:packet_dist_slope}. Although the volume of legitimate transmissions remains largely stable across all scenarios, the overall traffic increases substantially under attack conditions, reducing the proportional presence of normal packets within the dataset. The two distributions converge near the 50\% threshold at approximately 43 \textit{ms}, signifying the point at which adversarial activity begins to dominate the communication channel. Below this threshold, injected spam remains relatively stealthy, blending with benign BLE exchanges. Above it, spam traffic saturates the channel and becomes far easier to detect due to its overwhelming contribution to total packet volume.

From a defensive perspective, this crossover highlights the importance of adopting an adaptive duty-cycle strategy. Receivers equipped with the capability to dynamically regulate their scanning intervals can mitigate transient or low-frequency spam activity while maintaining efficient communication throughput. This agility allows for early-stage mitigation of low-impact interference while still ensuring timely detection and response once attacker behavior surpasses the observable threshold.

 While these experiments employed a single attacker at fixed delay intervals, the underlying model has broader implications. An agile adversary, capable of dynamically adjusting spam rates or transmitting in short bursts synchronized to observed scanning intervals, could achieve disproportionately high disruption with lower total packet volumes. In multi-attacker scenarios, even low-rate devices can collectively elevate the spam rate to disruptive levels without any single attacker drawing significant attention. For example, a single attacker at a moderate 100 ms rate achieves an estimated success probability of \~30\% for a moderate-duty-cycle receiver. Two such attackers operating independently elevate the success probability to roughly 40–60\%, while three attackers at the same rate reach 60–80\%, approaching the impact of a single aggressive 20 ms spammer.
 
 Overall, these empirical results reinforce the predictive value of the simulation, confirming that it accurately characterizes both brute-force disruption and more subtle, coordinated attacks. Moreover, the findings underscore that even moderately distributed adversaries can degrade BLE communication quality over time, posing significant challenges for systems that rely on duty-cycled or energy-constrained scanning strategies.

\section{Defense Against Persistent Spam Threats}
\label{sec:defense}

\subsection{Feasibility of Agility Based Defense: Multiple Agile Attackers Vs. Agile Victim}
In this agility-based defensive model, we introduce a dynamic and adaptive defense strategy to counteract the actions of multiple agile attackers. This model acknowledges the ever-changing nature of cyber threats and aims to create a resilient environment through agility. This approach assumes both attackers and victims to be agile. This defensive mechanism represents a more complex and dynamic environment compared to the previous models.

By using the same concept of agility (as in the attack model), if the defender can also dynamically and randomly change their parameters (spam rate (SR) for the attacker and the ratio duty cycle (DC) for the defender) over time, the resulting probability
calculation becomes more intricate. The dynamics of this defensive scenario and the interactions between the attacker and defender would need to be considered. We aim to study the feasibility of such a defensive posture by measuring the probability of a successful attack. 

A more general formulation will now involve time-dependent functions for both the attacker's spam rate $SR(t)$ and the defender's duty cycle $DC(t)$. The probability of a successful attack for an individual attacker $i^{th}$ at time \textit{t} could be represented as:

\begin{equation}
P_{\text{Spam}}^{i}(t) = \min\big(1, \max(0, DC(t) \times SR_i(t))\big), \quad \mapsto [0,1]
\label{eq:defense_individual_success}
\end{equation}

Therefore, the probability that no attacker succeeds at time t, updates to:

\begin{equation}
P_{\text{Spam}}^{\text{NS}}(t) = \prod_{i=1}^{N} \left(1 - P_{\text{Spam}}^{i}(t)\right)
\label{eq:defense_none_success}
\end{equation}

\begin{equation}
P_{\text{Spam}}^{\text{AOS}}(t) := 1 - P_{\text{Spam}}^{\text{NS}}(t) \equiv 1 - \prod_{i=1}^{N} \left(1 - P_{\text{Spam}}^{i}(t)\right), \quad \mapsto [0,1]
\label{eq:defense_any_success}
\end{equation}

Although, the final expressions for $P_{Spam}^{NS}(t)$ and $P_{Spam}^{AOS} (t)$ remain same (as in scenario 2), the impact of random variate $DC(t):\mapsto[0,1]$, over multiple BLE frequency channels, changes the outcome of the probability distribution. 

\begin{figure*}[!t]
\centering
\begin{minipage}[t]{0.495\textwidth}
    \centering
    \includegraphics[width=\linewidth]{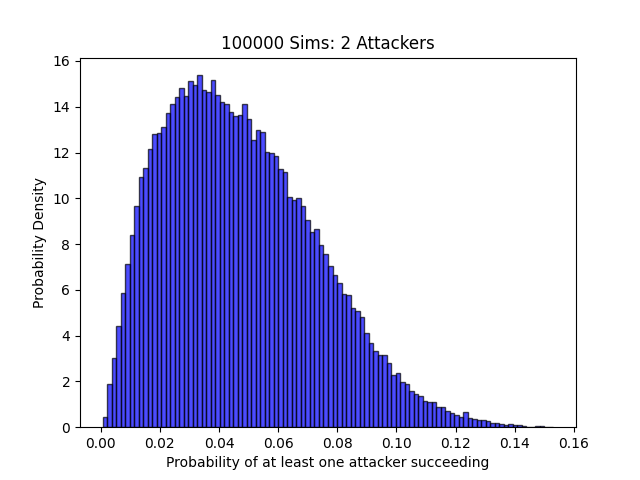}
\end{minipage}
\hfill
\begin{minipage}[t]{0.495\textwidth}
    \centering
    \includegraphics[width=\linewidth]{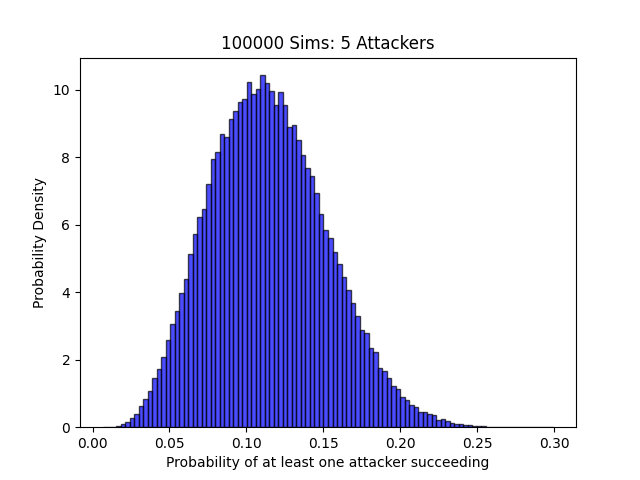}
\end{minipage}

\vspace{0.5em}

\begin{minipage}[t]{0.75\textwidth}
    \centering
    \includegraphics[width=\linewidth]{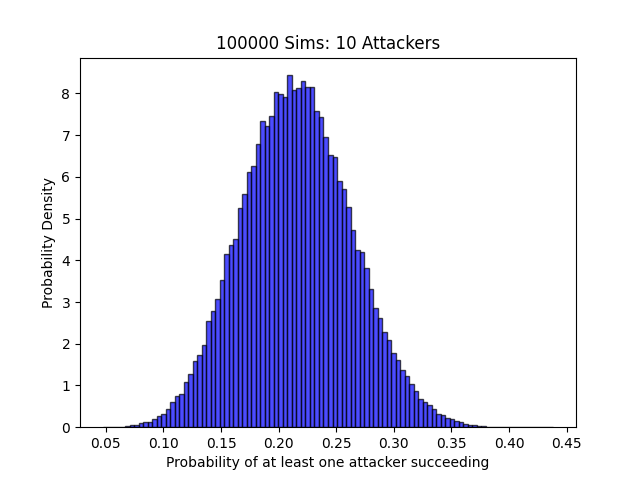}
\end{minipage}

\caption{Probability of at least one successful attack ($P_{\mathrm{Spam}}^{\mathrm{AOS}}(t)$) over time for varying numbers of agile attackers against an agile defender. Top-left: 2 attackers. Top-right: 5 attackers. Bottom: 10 attackers.}
\label{fig: agile probability distribution}
\end{figure*}

Figure~\ref{fig: agile probability distribution} illustrates the simulated probability distributions 
of $P_{\text{Spam}}^{\text{AOS}}(t)$ across 100,000 trials for 2, 5, and 
10-agile attackers facing an agile defender. With only two attackers, the 
distribution remains narrow and concentrated near low success probabilities, 
suggesting that an agile defender can reliably suppress the threat. As the 
number of attackers grows to five and then ten, the distribution 
progressively shifts toward higher success probabilities and exhibits 
greater spread, converging towards 22\% success rate for ten-attackers. 

A direct comparison between Figure~\ref{fig:10agile} and 
Figure~\ref{fig: agile probability distribution} (bottom panel) reveals the measurable advantage 
conferred by defender-side agility. In Figure~\ref{fig:10agile}, 
a static defender facing 10 agile attackers experiences near-certain 
disruption across virtually all combinations of spam rate and duty cycle, 
with $P_{\text{Spam}}^{\text{AOS}}(t)$ approaching $1.0$ across the 
majority of the parameter space. The contour plot leaves almost no safe 
operating region for the defender, even for lower duty cycles, confirming that a fixed duty cycle 
offers negligible resistance against a coordinated, adaptive 
multi-attacker threat.

In contrast, Figure~\ref{fig: agile probability distribution} (bottom 
panel) shows that when the defender also adopts agility by dynamically 
randomizing its duty cycle $DC(t)$, the maximum observed attack success 
probability converges to only $22\%$. This represents a substantial 
reduction compared to the near-certain disruption depicted in 
Figure~\ref{fig:10agile}, effectively multiplying the cost an attacker 
must bear to achieve the same level of disruption. The spread in the 
probability distribution further indicates that defender-side agility 
introduces genuine uncertainty into the attacker's ability to time spam 
transmissions successfully, even under sustained ten-attacker pressure, 
rendering coordinated flooding attempts considerably less predictable 
and harder to sustain.

\subsection{Design of Agility-based Defense Using First-Order Logic}
\label{sec: smt}
Since our main objective is to minimize the value of the probability of at least one successful attack, our objective function becomes:

\[Minimize (P_{Spam}^{AOS} (t)) \mapsto [0,1]\]

Although this can be solved by using optimization techniques (to find global minima), it is almost impossible to solve it in linear time, given the non-linear nature of the objective function. To tackle an agile attacker, this problem must be solved in near real-time, so that the defender can adjust its strategy on the go. Therefore, optimization might not be the ideal choice, hence opening the door for methods that may be able to find a counter-example in a certain scenario. For example, the optimization problem can be converted into a satisfiability problem, where we can add a threshold, and impose a constraint that the probability of attack must not exceed a certain acceptable threshold, instead of finding a global minimum value. 

In this case, we can formalize the problem using uninterpreted functions and variables in First-Order Logic as follows:
\begin{itemize}
\item Duty cycle of the defender:
$DC(t) \mapsto \{\, DC \in \mathbb{R} \mid 0 \leq DC(t) \leq 1 \,\}$.

\item Spam rate of attacker $i$:
$SR_i(t) \mapsto \{\, SR_i \in \mathbb{R} \mid 0 \leq SR_i(t) \leq 1,\; 1 \leq i \leq M \,\}$,
where $M$ denotes the maximum number of attackers.

\item Individual probability that attacker $i$ succeeds:
$P_{\mathrm{Spam}}^{i}(DC,SR) \mapsto
\{\, P_{\mathrm{Spam}}^{i} \in \mathbb{R} \mid 0 \leq P_{\mathrm{Spam}}^{i}(t) \leq 1 \,\}$.
\end{itemize}


\vspace{0.8em}
\noindent\textbf{Combined success probability.}
Let $M\in\mathbb{N}_{\ge 1}$ be the maximum number of attackers defined
above.
Assuming independent attempts, the probability that \emph{at least one}
attacker succeeds at time~$t$ is
\[
\boxed{
P_{\mathrm{Spam}}^{\mathrm{AOS}}(t)
=
1-\prod_{i=1}^{M}
\Bigl(1-P_{\mathrm{Spam}}^{i}\bigl(DC(t),SR_i(t)\bigr)\Bigr)
}
\]

When the separable model
$P_{Spam}^{i}(d,s)=d\times s$ is instantiated, this reduces to
$P_{Spam}^{AOS}(t)=1-\bigl(1-DC(t)\,SR_1(t)\bigr)\cdots
                       \bigl(1-DC(t)\,SR_M(t)\bigr)$,
which is the expression implemented in the prototype SMT script.

\medskip
\noindent\textbf{Threshold formulation.}
Rather than minimising $P_{Spam}^{AOS}(t)$ outright, we select an
acceptable upper bound $T\in[0,1]$ (e.g.\ $T=0.35$) and restrict the
\emph{search space} of duty--cycle and spam‑rate parameters to a
defender‑defined operating window
\[
\boxed{
\begin{gathered}
\underline{D} < DC(t) < \overline{D},\\
\underline{S} < SR_i(t) < \overline{S},\quad 1 \le i \le M
\end{gathered}
}
\]

where $\underline{D},\overline{D},\underline{S},\overline{S}$ are
constant bounds supplied by policy.

\medskip
\noindent\textbf{Existential counter‑example query (SMT).}
The optimization problem is recast as the following satisfiability
question:
\[
\boxed{
\begin{gathered}
\exists\, (DC,SR):\;
\underline{D} < DC(t) < \overline{D} \\
\land\; \bigwedge_{i=1}^{M}
\bigl(\underline{S} < SR_i(t) < \overline{S}\bigr) \\
\land\; P_{\mathrm{Spam}}^{\mathrm{AOS}}(t) > T
\end{gathered}
}
\]
If the SMT solver returns \textsc{sat}, it provides concrete values
$DC(t)$ and $SR_i(t)$ that violate the threshold, signaling the
defender to adapt its parameters (and avoid unsafe configuration).  A result of \textsc{unsat} certifies
that no admissible configuration exceeds~$T$ for the current~$M$.

\begin{figure*}[h]
    \centering
    \includegraphics[width=0.95\textwidth]{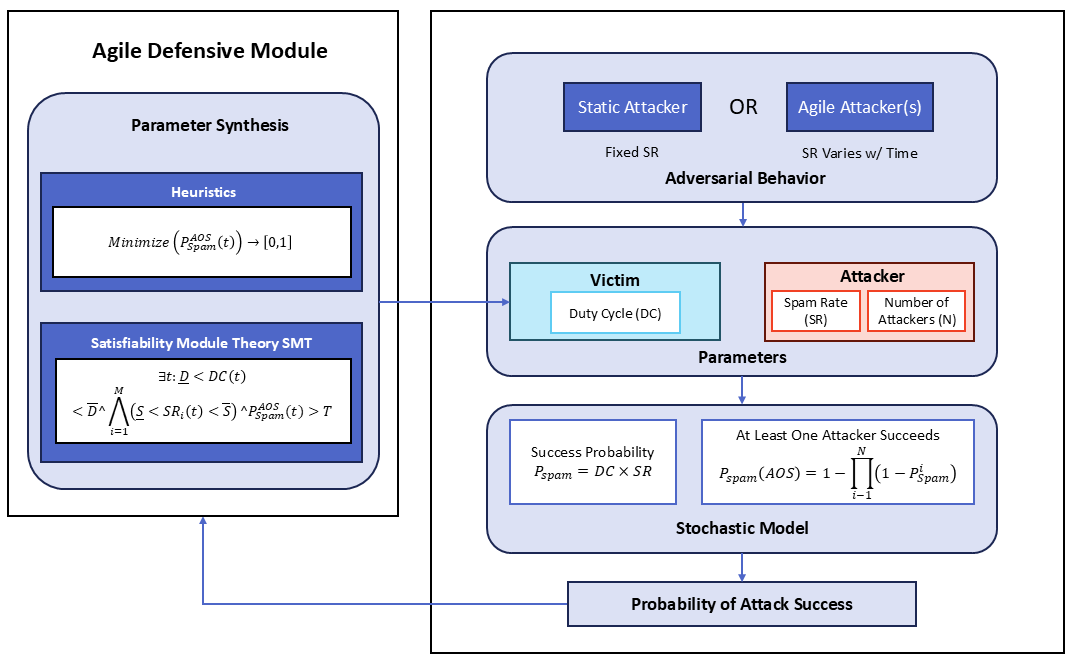}
    \caption{Conceptual structure of the agility-based defense framework. The defender employs heuristic optimization and SMT-based constraint reasoning to synthesize adaptive duty-cycle parameters, while the adversary may operate as a static or agile attacker with time-varying spam rates. }

    \label{fig:Defense_fig}
\end{figure*}

Figure~\ref{fig:Defense_fig} presents the conceptual architecture 
of the proposed agility-based defense framework. The adversarial side 
of the model accommodates both static and agile attackers: a static 
attacker operates with a fixed spam rate $SR$, while an agile attacker 
varies its transmission behavior over time, represented by the 
time-dependent function $SR_i(t)$. The defense module, relies on the stochastic model to calculate the probabilities, and correspondingly utilizes our above mentioned formalization and  Satisfiability Modulo Theories (SMT) solver, to verify, whether a safe state exists for certain parameters. This safety information is then relayed to the defender to synthesize strategy, such that the corresponding parameters can be tuned to avoid any further jamming, spam, or DoS attacks.

\section{Conclusion \& FutureWork}
\label{sec:conclusion}

This work demonstrate the measurable impact of BLE spam and flooding attacks on IoT environments and establishes a foundation for understanding their disruptive capacity. We formalized the the problem as a stochastic process and presented: (i) precise threat models of both static and adaptive adversaries and their interaction with IoT environment; (ii) empirical validation through a controlled testbed that captured real-world BLE traffic across varying levels of attacker aggression; and (iii) integration of defender-side agility into the same stochastic framework to evaluate response strategies under dynamic conditions. To the best of our knowledge, this is the first effort to model adversarial behavior in IoT environments, combining empirical validation with formal analytical tools to characterize attacker capabilities and inform actionable defense strategies.

The findings underscore two key insights. First, even moderately aggressive attackers, when distributed across multiple devices, can achieve levels of disruption comparable to a single, high-intensity adversary. Second, defender-side agility can reduce the probability of attack success below critical/desired  thresholds. However, translating this potential into practice remains a significant challenge. By jointly modeling attacker adaptability and system responsiveness, our study offers BLE implementers a quantitative basis to anticipate degradation risks and reason about adaptive mitigation strategies.

Through this adversarial behavioral modeling, our study contributed both validated empirical evidence and actionable insight. The combination of probabilistic modeling and testbed validation not only quantified the impact of stealthy BLE flooding attacks but also produced design-level recommendations for strengthening IoT deployments. While our defense-side modeling provides a promising foundation, future work should extend these results into deployable countermeasures and policy-level guidance for standardization bodies such as the Bluetooth SIG.

In particular, future work will explore the impact and practical viability of channel hopping as a defensive strategy, including the extent to which BLE stack updates or modifications can support more agile channel-selection behavior. We also aim to extend the proposed stochastic model to explicitly study multi-channel effects, enabling a more realistic analysis of adversarial flooding and resilience emerging from defender's adaptation across multiple BLE advertising channels. Finally, we plan to collect a more rigorous and diverse dataset to further validate the defensive capabilities of the proposed model under realistic Internet of Medical Things (IoMT) deployment conditions.

\section*{Datasets}
All the data used in this paper can be made available upon request.

\bibliography{references}

\end{document}